\documentclass[prc,twocolumn,showpacs,superscriptaddress,amsmath,amssymb]{revtex4}

\usepackage{graphicx}
\usepackage{dcolumn}
\usepackage{bm}

\usepackage[none]{hyphenat}

\begin{document}

\title{Measurements of $^{152}$Gd(p,$\gamma$)$^{153}$Tb and $^{152}$Gd(p,n)$^{152}$Tb reaction cross sections
for the astrophysical $\gamma$ process}

\author{R. T. G\"{u}ray\footnote{Corresponding Author: tguray@kocaeli.edu.tr }
}%
\author{N. \"{O}zkan}
\author{C. Yal\c{c}{\i}n}
\affiliation{Department of Physics, Kocaeli University, Umuttepe
41380, Kocaeli, Turkey
}%
\author{T. Rauscher}
\affiliation{Centre for Astrophysics Research, School of Physics, Astronomy and Mathematics, University of Hertfordshire, Hatfield 9AL 10AB, United Kingdom}
\affiliation{Department of Physics, University of Basel,
CH-4056 Basel, Switzerland
}%

\author{\mbox{Gy. Gy\"{u}rky}}
\author{J. Farkas}
\author{Zs. F\"{u}l\"{o}p}
\author{Z. Hal\'{a}sz}
\author{E. Somorjai}
\affiliation{MTA Institute for Nuclear Research (MTA Atomki), 4001 Debrecen, Hungary}%

\date{\today}

\begin{abstract}

The total cross sections for the $^{152}$Gd(p,$\gamma$)$^{153}$Tb 
and $^{152}$Gd(p,n)$^{152}$Tb reactions have been measured by the
activation method at effective center-of-mass energies \mbox{$3.47
\leq E_\mathrm{c.m.}^\mathrm{eff}\leq 7.94$ MeV} and \mbox{$4.96
\leq E_\mathrm{c.m.}^\mathrm{eff} \leq 7.94$ MeV}, respectively.
The targets were prepared by evaporation of 30.6\%
isotopically enriched $^{152}$Gd oxide on aluminum backing
foils, and bombarded with proton beams provided by a cyclotron 
accelerator. The cross sections were deduced from the observed $\gamma$-ray activity,
which was detected off-line by a HPGe detector in a low background environment. 
The results are presented and compared with
predictions of statistical model calculations. This comparison supports a modified optical proton+$^{152}$Gd potential suggested earlier. 
\end{abstract}


\maketitle

\section{Introduction}
\label{sec:intro}

Only the lightest elements (hydrogen, helium, and traces of lithium, beryllium) originate from a hot Big Bang phase in the early stage of the Universe \cite{steigman,winteler,rauinhom}. All other nuclei have been produced thereafter in stars and stellar explosions. Most nuclides above the Fe group (mass numbers \mbox{$A\gtrsim 60$}) originate from neutron captures and subsequent $\beta$ decays and thus are called neutron-capture elements \cite{b2fh,cameron,mey94}. Two main processes have been identified, the $s$ process (slow neutron capture process, with a main component occurring in medium mass Asymptotic Giant Branch (AGB) stars and a weak component stemming from massive stars) and the $r$ process (rapid neutron capture process) \cite{Kappeler90,Arnould07,ctt}. The site of the latter process is disputed but core-collapse supernova explosions and/or neutron star mergers may provide viable alternatives \cite{arconesthiele}.
There are a number of proton-rich isotopes between Se and Hg which cannot be produced by either process. They are called $p$ nuclei and further nucleosynthesis processes have to be invoked to explain their existence \cite{p-review}. Among the originally 35 $p$ nuclei \cite{b2fh,cameron}, large $s$ process contributions to $^{164}$Er, $^{152}$Gd, and $^{180}$Ta were found in modern AGB models \cite{arl99}. Also the abundances of $^{113}$In and $^{115}$Sn may receive contributions from the $s$ and/or $r$ process \cite{nemeth}.

The currently accepted production mechanism for the bulk of $p$ nuclei is the so-called $\gamma$ process, synthesizing proton-rich isotopes by photodisintegration reactions \cite{arnould,p-review}. The $\gamma$ process occurs in the outer layers of a massive star during its core-collapse supernova explosion \cite{Woosley,Rayet95,Rauscher02}. As an additional site, type Ia supernovae have been suggested, where also the required temperatures are achieved \cite{howmey,kusa11,travWD,trav14}. In both sites, quickly expanding zones of hot matter show production of $p$ nuclei when the reached peak temperatures are in the range \mbox{$2\leq T\leq 3.5$ GK}. In order to produce the heavier $p$ nuclei it is necessary to stay at the lower end of this temperature range because they would be completely destroyed at higher temperatures. On the other hand, the higher end of the temperature range is required for the production of the light $p$ nuclei because lighter nuclei are more strongly bound and less easily photodisintegrated \cite{p-review,raubranch,Rapp06}.

The very rare $^{138}$La cannot be produced in a $\gamma$ process but it was suggested to be formed through neutrino reactions on stable nuclei in core-collapse supernovae, the so-called $\nu$ process \cite{neutrino,hekol05}.

A $\gamma$ process starts with ($\gamma$,n) reactions on stable nuclei already present in the stellar plasma. When the temperatures remain high for a sufficiently long time, proton-rich isotopes are reached for which ($\gamma$,p) or ($\gamma$,$\alpha$) reactions are faster than ($\gamma$,n). This leads to a deflection of or branching in the nucleosynthesis path \cite{raubranch,Rapp06}. Due to the nuclear structure influencing the binding energies, ($\gamma$,p) deflections are mainly found in nuclei with neutron number \mbox{$N<82$}, whereas ($\gamma$,$\alpha$) branchings are important in the region \mbox{$N\geq 82$} \cite{p-review,raubranch}.

\textit{Stellar} photodisintegration \textit{rates} used in reaction network calculations are usually derived from \textit{stellar} capture \textit{rates} by applying the principle of detailed balance \cite{fow76,holmes,woosleyadndt,Raus00,raureview}. It should be noted that this reciprocity only applies to \textit{stellar} rates, involving thermally excited nuclei in entrance and exit channel, but not to laboratory capture cross sections and their respective photodisintegration cross sections, unless the cross sections are dominated by single transitions between initial state, compound state, and final state \cite{fow76,Ili07,blatt}. Reactions involving light nuclei often include only a few transitions but this is never the case in the regime of high level-densities encountered in nuclei participating in the synthesis of $p$ nuclei. Although direct ($\gamma$,n) measurements have been performed also for intermediate and heavy nuclei (e.g., \cite{sprocess_w,Utsunomiya06,Mohr07}), it has been shown that these cannot constrain stellar rates, as they only allow to study a single $\gamma$ transition from the ground state of the target nucleus. Its contribution to the stellar rate is less than 0.1\%, contrary to capture data which allow to constrain a much larger fraction (on the order of several tens of percent, depending on plasma temperature and the nuclear level structure) of the stellar capture rate (see, e.g., \cite{p-review,raureview,Mohr2,Kiss08,Rau09,raugamma,rausensi,rauAIP}). The contribution of ($\gamma$,p) and ($\gamma$,$\alpha$) ground-state transitions to the stellar rate is even lower due to the shift of the Gamow window to higher compound excitation energies \cite{rausensi}. Interestingly, the ground-state contribution  of the respective capture rates even is larger than those for (n,$\gamma$) at the same temperature due to Coulomb suppression of transitions with low relative energy \cite{Kiss08,Rau09,raucoulsupp}. This even applies to captures with negative reaction $Q$ values on intermediate and heavy targets, allowing ground-state capture to always constrain a much larger fraction of the stellar rates than photodissociation measurements.

Although capture measurements still cannot completely constrain a high-temperature stellar rate, either, they can be used to test the predictions of charged particle widths. Proton and $\alpha$ widths determine the stellar photodisintegration rates because $\gamma$-process temperatures,
although being high even for stellar plasmas, still translate to comparatively low nuclear interaction energies, below the Coulomb barrier when involving charged particles. This results in charged particle widths being smaller than the $\gamma$ widths and thus a dependence of the rates primarily on these charged particle widths rather than on the $\gamma$ widths \cite{p-review,rausensi}. Experiments can probe the charged particle widths in capture reactions but the small cross sections at subCoulomb energies make it difficult to measure in the astrophysical energy range. Therefore, astrophysically relevant data are scarce. The few available data have shown considerable deviations from predictions, especially for reactions with $\alpha$ particles \cite{p-review,Som98,sm144lett}. Low-energy proton captures have been found to be predicted more reliably, although small modifications to the optical proton+nucleus potentials have been suggested \cite{Kiss07}.

Proton capture reaction cross sections for the $\gamma$ process have been measured before, e.g., by \cite{{Kiss07}, Guray09, Gyu03, Gyu01, Galan03, Tsagari04, Laird87, Har01, Sauter97, Chloupek99, Bork98, Ozkan01, Ozkan02, Gyu07, Famiano08, Kiss08, Spy08, Sauerwein12, Net14}. Since (p,$\gamma$) above the (p,n) threshold are sensitive not only to the proton width but also to the neutron and $\gamma$ widths, additional information is required to be able to distinguish the impact of the various widths (see also \mbox{Sec.\ \ref{sec:sensi}}). Such information can be obtained through a simultaneous (p,n) measurement because the (p,n) cross section well above the threshold is mainly determined by the proton width.
Some (p,n) measurements have been performed on $s$ or $r$ process seed nuclei that affect the abundances of the
light $p$ nuclei, e.g., 
$^{76}$Ge \cite{Kiss07}, $^{82}$Se \cite{Gyu03}, $^{85}$Rb
\cite{Kiss08}, and on a $p$ nucleus $^{120}$Te \cite{Guray09}.

Here, we present a simultaneous measurement of (p,$\gamma$) and (p,n) cross sections of $^{152}$Gd. Although the $\gamma$-process contribution to $^{152}$Gd is small and originates from ($\gamma$,n) reactions on other Gd isotopes, a determination of proton-induced cross sections at low energy of this proton-rich, heavy nucleus allows to test statistical model predictions well below the Coulomb barrier. For the first time, it was also possible to determine proton capture reaction cross sections below the (p,n) threshold for this nucleus.

The $^{152}$Gd(p,$\gamma$)$^{153}$Tb and
$^{152}$Gd(p,n)$^{152}$Tb activation measurements have
been performed in an energy range between 3.47 and \mbox{7.94 MeV} as a test of the statistical
model predictions over a broader energy region. The details of
the experiment are given in \mbox{Sec.\ \ref{sec:exp}}. The experimental
cross sections are compared to predictions in the
Hauser-Feshbach statistical model in \mbox{Sec.\ \ref{sec:results}}. 
A summary and conclusions are provided in \mbox{Sec.\ \ref{sec:summary}}.

\begin{table*} 
\caption{\label{tab:decay}Decay parameters of the p+$^{152}$Gd reaction
products \cite{nudat} and measured photo-peak efficiencies at energies of the
$\gamma$ transitions, as used in the analysis.}
\setlength{\extrarowheight}{0.1cm}
\begin{ruledtabular}
\begin{tabular}{cccccc}
\parbox[t]{2.0cm}{\centering{Reaction}} &
\parbox[t]{1.5cm}{\centering{Product}} &
\parbox[t]{2.0cm}{\centering{Half-life }} &
\parbox[t]{2.0cm}{\centering{$\gamma$ Energy (keV)}} &
\parbox[t]{2.0cm}{\centering{$\gamma$ Intensity (\%)}}&
\parbox[t]{3.0cm}{\centering{  Detection efficiency (\%)}} \\
\hline
$^{152}$Gd($p$,$\gamma)$& $^{153}$Tb & (2.34 $\pm$ 0.01) d & 212.00 &31.0 $\pm$ 0.2& 1.83 $\pm$ 0.09\\
$^{152}$Gd($p,n$)& $^{152g}$Tb & (17.5 $\pm$ 0.1) h & 271.08 & 8.6 $\pm$ 0.8 & 1.53 $\pm$ 0.08 \\
~~~~~~~~~~~&                   &                    & 344.28 & 65.0 $\pm$ 0.1 & 1.27 $\pm$ 0.06 \\
\end{tabular}
\end{ruledtabular}
\end{table*}

\section{Experimental Procedure}
\label{sec:exp}

The reaction cross sections of $^{152}$Gd(p,$\gamma$)$^{153}$Tb and $^{152}$Gd(p,n)$^{152}$Tb 
were measured using an activation technique. The target preparation and the experiment were performed 
using the laboratory facilities at the Institute for Nuclear Research of the Hungarian Academy of Sciences 
(MTA Atomki). After the target preparation, the experiment has two stages, 
first the activation of the target and then   
the determination of the number of \mbox{$\beta$-unstable} reaction products using an \mbox{off-line} gamma detection system. The measurement of proton induced reactions with the activation technique  is discussed in detail in Refs.\ \cite{{Ozkan02}, {Famiano08}}.

\subsection{Target preparation}

The 30.6\% isotopically enriched $^{152}$Gd oxide in powder form was provided by the company ISOFLEX USA, Certificate No: 64-02-152-1453. For the target backing, high purity thin \mbox{(2.5 $\mu$m)} aluminum foils, obtained from Espi Metal Ltd, were used. The targets were prepared by vacuum evaporation. 

There are two methods available in the vacuum chamber to evaporate the target material; via the resistive heating method and the use of an electron gun. Because of the high melting point \mbox{($\approx 2350$ $^\circ$C)} of gadolinium oxide Gd$_{2}$O$_{3}$, the electron gun was used for the evaporation. Before the evaporation, \mbox{40 mg} Gd$_{2}$O$_{3}$ powder was pressed into a pellet with \mbox{6 mm} diameter. This Gd$_{2}$O$_{3}$ pellet was placed in a tantalum boat and directly heated by the electron beam. 

The target holder was made of aluminum with 9 holes (each hole has \mbox{1.2 cm} inside diameter) and Al backing foils with the size of \mbox{$1.5\times 1.5$ cm$^{2}$} were placed on these holes. The target holder was placed \mbox{7 cm} above the tantalum boat for Gd$_{2}$O$_{3}$ deposition. At the same time 9 targets were produced with thicknesses varying between \mbox{220 $\mu$g/cm$^{2}$} and \mbox{310 $\mu$g/cm$^{2}$}. After the evaporation, the targets were fixed into target frames made of aluminum with \mbox{3.8 cm} outer diameter, \mbox{1.2 cm} inner diameter, and \mbox{3 mm} thickness. The target thickness was determined by weighing. The weight of the Al foil was measured before and after the evaporation with a precision better than \mbox{5 $\mu$g} and  the Gd$_{2}$O$_{3}$ number density could be determined from the difference. The target with the thickness of \mbox{310 $\mu$g/cm$^{2}$} was also examined by the Rutherford Back Scattering (RBS) method with a microprobe to investigate the target homogeneity. The RBS spectra were taken at the Van de Graaff accelerator of MTA Atomki with \mbox{2.0 MeV} $\alpha$ particles using a \mbox{$3 \times 3$ $\mu$m$^{2}$} beam spot size and \mbox{$100 \times 100$ $\mu$m$^{2}$} scanning size. The largest difference in the target thickness between two points on the target was found to be 7\%.

\subsection{Activation}

\begin{figure}
\includegraphics[width=\columnwidth]{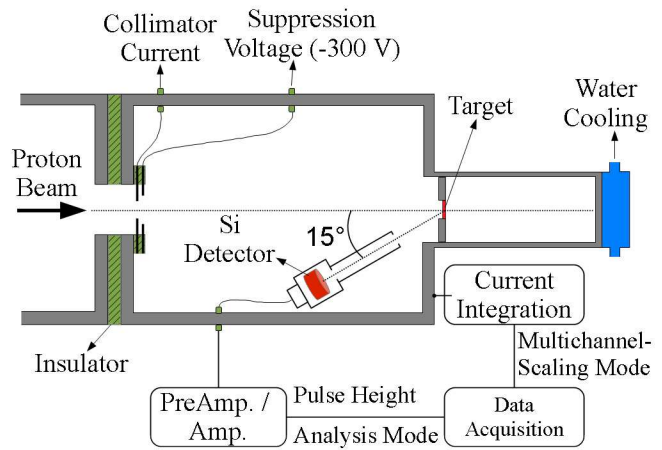}
\caption{\label{fig:schematic}(Color online) A schematic diagram of the irradiation chamber mounted
at the end of the beam line at the cyclotron accelerator of MTA Atomki.}
\end{figure}

In order to activate the $^{152}$Gd targets, they were irradiated with
a proton beam in an energy range between 3.5  and \mbox{8.0 MeV} at the 
cyclotron accelerator of MTA Atomki. The proton energy was changed by 
\mbox{0.5 MeV} steps in the laboratory frame. 
Figure \ref{fig:schematic} shows a schematic diagram of 
the irradiation chamber mounted at the end of the beam line. The beam current 
was measured with a current integrator using the entire chamber as 
a Faraday cup, after the last beam defining aperture. The beam current was kept  
as stable as possible at \mbox{2 $\mu$A}. The integrated current was recorded
in time intervals of one minute and the changes in the beam current during the 
activation was taken into account in the data analysis. The activations for different 
proton energies lasted between 0.5 and 30 hours, based on the requirement of sufficient counting statistics. 

In order to suppress the secondary electrons from the targets, a bias voltage of
\mbox{$-300$ V} was applied to the entrance of the target irradiation chamber.
Backscattered protons were detected in order to monitor the target
stability during each irradiation by using an ion implanted Si detector 
with a reduced entrance aperture of \mbox{0.5 mm}
diameter at an angle of \mbox{165$^\circ$} with respect to
the beam direction, as shown in Fig.\ \ref{fig:schematic}.

The target taken from the irradiation chamber after the activation was 
transferred to the \mbox{off-line} gamma detection system to determine 
the number of reaction products of the proton induced reactions on $^{152}$Gd.

\subsection{Determination of the $^{153}$Tb and $^{152}$Tb activities}
\label{sec:analysis}

\begin{figure*} 
\includegraphics[width=\textwidth]{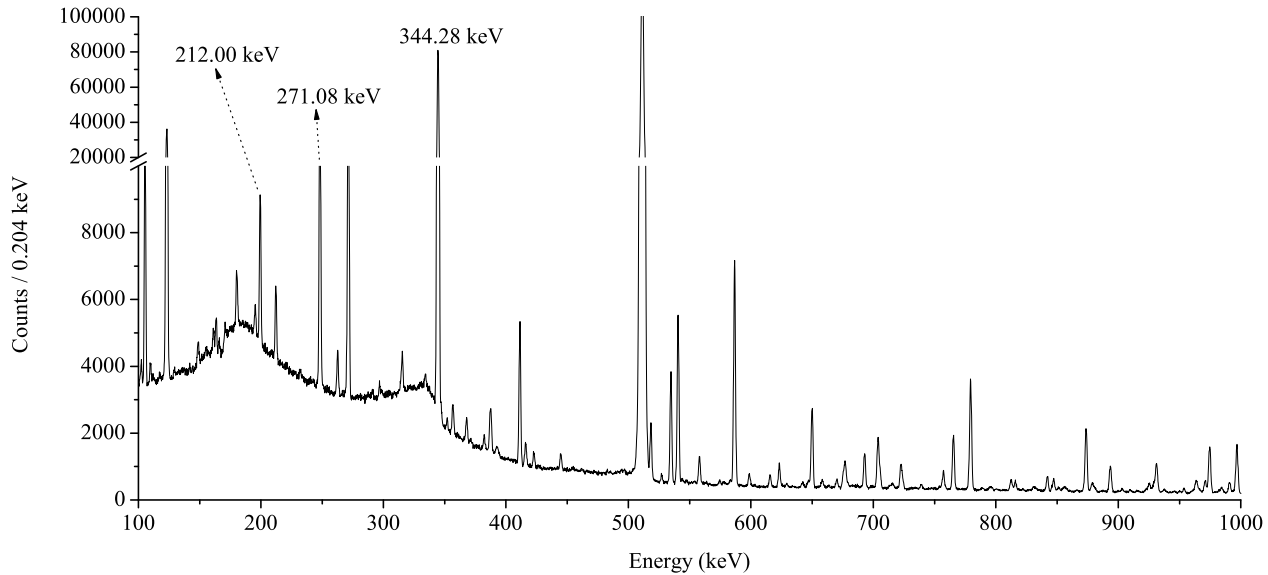}
\caption{\label{fig:spectrum}The $\gamma$-ray spectrum recorded for a counting time of 3.64 hours 
after 2.08 hours of activation with a \mbox{7 MeV} proton beam.
The $\gamma$ transitions used in the measurements are indicated by arrows. }
\end{figure*}

In order to determine the number of Tb isotopes, the $\gamma$ radiation following 
their electron-capture decays was counted with a 100\% relative efficiency 
HPGe detector in a complete 4$\pi$ low background lead shielding. 
After the end of each activation, the $^{152}$Gd target was mounted in a holder placed \mbox{10 cm} 
from the end of the detector cap. The $\gamma$ spectra were taken between 
2.5 and 114 hours, depending on the counting statistics, and were stored regularly in every hour, 
thus different isotope decays could be analyzed.

The activation technique can be used to measure 
the reaction cross sections of $^{152}$Gd(p,$\gamma$)$^{153}$Tb and 
$^{152}$Gd(p,n)$^{152}$Tb because $\beta$-decay half-lives of the reaction 
products in both cases are long enough to determine their yields (Table \ref{tab:decay}). For
$^{152}$Gd(p,n)$^{152}$Tb, the product $^{152}$Tb has ground
($^{152g}$Tb) and isomeric states ($^{152m}$Tb), but $^{152m}$Tb disintegrates 
78.8\% through internal transitions and 21.2\% by $\beta$ decay \cite{nudat}. The contribution of $^{152m}$Tb to 
the total $^{152}$Gd(p,n)$^{152}$Tb reaction cross section is negligible within uncertainties, as was verified in a calculation with the TALYS code \cite{Talys01}.

For the analysis of the $^{152}$Gd(p,$\gamma$)$^{153}$Tb reaction data, the \mbox{212.00 keV} $\gamma$ 
line was used, which has an emission probability of 31.0\% in electron-capture of $^{153}$Tb 
(half-life \mbox{2.34 d}). In the case of the $^{152}$Gd(p,n)$^{152}$Tb 
reaction, 271.08 and {344.28 keV} $\gamma$ lines with emission probabilities of 8.6\% and 65.0\%, respectively, 
were used because they are associated with the decay of the ground state of $^{152}$Tb with a half-life of \mbox{17.5 h}. 
The $\gamma$ transitions and decay parameters used for the data analysis are summarized in Table \ref{tab:decay}.
Figure \ref{fig:spectrum} shows the $\gamma$-ray spectrum recorded for a counting time of \mbox{3.64 h} 
after a \mbox{2.08 h} activation with a \mbox{7 MeV} proton beam. 
The $\gamma$ transitions used in the measurements, as given in Table \ref{tab:decay}, are also indicated in the figure. 

The $\gamma$-detection system was the same as the one used in a previous $^{130, 132}$Ba experiment 
and is described in details in \cite{Hala12}. The absolute photopeak efficiency calibration
was performed for a detector-target distance of 27 cm, where true coincidence summing effects are negligible, using the calibrated sources, $^7$Be, $^{22}$Na, $^{54}$Mn, $^{57}$Co, $^{60}$Co, 
$^{65}$Zn, $^{133}$Ba, $^{137}$Cs, $^{152}$Eu, and $^{241}$Am. Although the true coincidence 
summing effect could be expected to be small at a distance of more than 10 cm between the detector and the target, 
it has been taken into account using the following method. 

A $^{152}$Gd target was irradiated at a proton
energy of \mbox{8.0 MeV} in the laboratory frame and then the $\gamma$ spectrum of the reaction products 
was measured at both 27 and \mbox{10 cm} distances.
Taking into account the time elapsed between the two countings, 
the ratio of the count rates taken at 10 cm to 
the one taken at \mbox{27 cm} gives the ratio of the photo-peak efficiency at \mbox{10 cm} to one at \mbox{27 cm}. 
In this way, an efficiency value at \mbox{27 cm} can be normalized to the one at \mbox{10 cm} 
covering the coincidence effect correction for the $\gamma$ transition used for the analysis.
The photo-peak efficiencies of the $\gamma$ transitions used to identify the products of the investigated reactions are also given in Table \ref{tab:decay}.

\section{Results and discussion}
\label{sec:results}

\subsection{Experimental cross sections}

\begin{figure}
\includegraphics[width=\columnwidth]{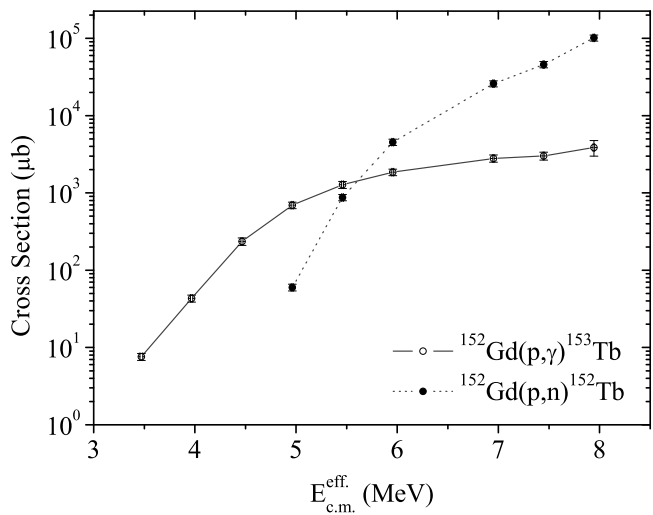}
\caption{\label{fig:exp_xsec} Experimental cross sections of $^{152}$Gd(p,$\gamma$)$^{153}$Tb 
and $^{152}$Gd(p,n)$^{152}$Tb reactions.}
\end{figure}

The cross sections of $^{152}$Gd(p,$\gamma$)$^{153}$Tb and
$^{152}$Gd(p,n)$^{152}$Tb have been measured at proton 
energies ranging from 3.5 to \mbox{8.0 MeV}, in steps of \mbox{0.5 MeV}. 
This energy range corresponds to effective center-of-mass
energies between 3.47 and \mbox{7.94 MeV}. The effective
center-of-mass energies $E_\mathrm{c.m.}^\mathrm{eff}$ are defined as the 
center-of-mass energies at which one half of the reaction yield
for the entire target thickness is obtained \cite{{Ili07},
{Rofls88}}. 
The experimental cross section results for the (p,$\gamma$) and (p,n) reactions are summarized 
in Tables \ref{tab:pg} and \ref{tab:pn}, respectively, and are also shown in \mbox{Fig.\ \ref{fig:exp_xsec}}.
The (p,n) channel becomes dominant very fast above its threshold energy \mbox{(4.8 MeV)} and 
its cross section becomes much higher than that of the $^{152}$Gd(p,$\gamma$)$^{153}$Tb reaction 
\mbox{(Fig.\ \ref{fig:exp_xsec})}.

\begin{table}
\caption{\label{tab:pg}Measured cross sections of the
$^{152}$Gd(p,$\gamma$)$^{153}$Tb reaction.}
\begin{ruledtabular}

\begin{tabular}{ccc}

$E_\mathrm{beam}$ & $E_\mathrm{c.m.}^\mathrm{eff}$ & Cross section \\

(MeV) &  (MeV) & ($\mu$b) \\

\hline

3.500 & 3.471 $\pm$ 0.014 &   7.6 $\pm$ 0.8\\
4.000 & 3.968 $\pm$ 0.012 &   43.2 $\pm$ 4.4 \\
4.500 & 4.466 $\pm$ 0.010 &   235 $\pm$ 26 \\
5.000 & 4.962 $\pm$ 0.011 &   692 $\pm$ 65 \\
5.500 & 5.460 $\pm$ 0.009 &   1277 $\pm$ 133  \\
6.000 & 5.956 $\pm$ 0.009 &   1850 $\pm$ 178  \\
7.000 & 6.951 $\pm$ 0.007 &   2796 $\pm$ 298\\
7.500 & 7.447 $\pm$ 0.006 &   3013 $\pm$ 356 \\
8.000 & 7.943 $\pm$ 0.008 &   3875 $\pm$ 887  \\
\end{tabular}
\end{ruledtabular}
\end{table}

\begin{table}
\caption{\label{tab:pn}Measured cross sections of the
$^{152}$Gd(p,n)$^{152}$Tb reaction.}
\begin{ruledtabular}

\begin{tabular}{ccc}
$E_\mathrm{beam}$ & $E_\mathrm{c.m.}^\mathrm{eff}$ & Cross section  \\
(MeV)         & (MeV)                          & ($\mu$b) \\ 

\hline 
5.000 & 4.962 $\pm$ 0.011 & 60 $\pm$ 6  \\
5.500 & 5.460 $\pm$ 0.009 & 875 $\pm$ 81  \\
6.000 & 5.956 $\pm$ 0.009 & 4530 $\pm$ 413  \\
7.000 & 6.951 $\pm$ 0.007 & 25915 $\pm$ 2365 \\
7.500 & 7.447 $\pm$ 0.006 & 45739 $\pm$ 4174 \\
8.000 & 7.943 $\pm$ 0.008 & 101314 $\pm$ 9241  \\

\end{tabular}
\end{ruledtabular}
\end{table}

As mentioned in \mbox{Sec.\ \ref{sec:intro}}, the product of $^{152}$Gd(p,n) reaction has both a ground
($^{152g}$Tb) and an isomeric state ($^{152m}$Tb). Based on the results of a TALYS calculation \cite{Talys01}, 
the contribution of $^{152m}$Tb to the (p,n) reaction cross sections is less than 0.9\%. This contribution can be neglected, especially because the short-lived isomer decays by high probability (78.8\%) to the long-lived ground state of $^{153}$Tb, and is therefore included in the counting of the ground state decay.  
For the (p,n) reaction, hence, measurements of only $^{152}$Gd(p,n)$^{152g}$Tb have been performed.
The measured (p,n) cross sections determined from two different $\gamma$ transitions, 
as explained in \mbox{Sec.\ \ref{sec:analysis}}, were found to be statistically consistent with each other.
As a result, the weighted averages of the $^{152}$Gd(p,n)$^{152}$Tb reaction
cross sections deduced from these two $\gamma$ transitions are presented.

The uncertainty in the final results has been determined based on the propagation of partial
errors: counting statistics (0.2\% to 21\%),  
decay parameters (0.4\% to 9\%), detection efficiency (5\%), 
target thickness ($\sim$ 7\%), and beam current integration (less than 3\%).
The uncertainties in the effective center-of-mass
energies range between 0.08\% and 0.4\%; they were
calculated with the SRIM code \cite{SRIM} based on the proton
energy loss in the targets.

\subsection{Comparison with Hauser-Feshbach statistical model predictions} 
\label{sec:theory}

\subsubsection{Sensitivity of reaction cross sections to modifications in the predicted total widths}
\label{sec:sensi}

In order to gauge the dependence of the cross sections on the various ingredients of the calculation and to understand what properties can be constrained by a comparison of predictions and experiment, it is useful to first consider the sensitivities of the astrophysical reaction rates and cross sections. The sensitivity $s$ of a cross section or rate $\mathcal{C}$ is extensively discussed in \cite{rausensi}. It is defined as
\begin{equation}
s=\frac{\frac{\mathcal{C}_\mathrm{new}-\mathcal{C}_\mathrm{old}}{\mathcal{C}_\mathrm{old}}}{\frac{W_\mathrm{new}-W_\mathrm{old}}{W_\mathrm{old}}} \quad,
\end{equation}
as is the standard in general sensitivity analysis.
In the current context, the sensitivity measures by how much a rate or cross section $\mathcal{C}$ changes when one of the averaged total widths $W$ appearing in the Hauser-Feshbach formalism is varied. For example, if the cross section changes by the same factor by which a width is varied, the sensitivity is \mbox{$s=1.0$}, if it does not change, \mbox{$s=0$}. The sign of $s$ provides information on whether the change in $\mathcal{C}$ is in the same direction as the change in the width \mbox{($s>0$)}, i.e., increase with increasing width, or in opposite direction \mbox{($s<0$)}, i.e., decrease with increasing width.

\begin{figure}
\includegraphics[angle=-90,width=\columnwidth]{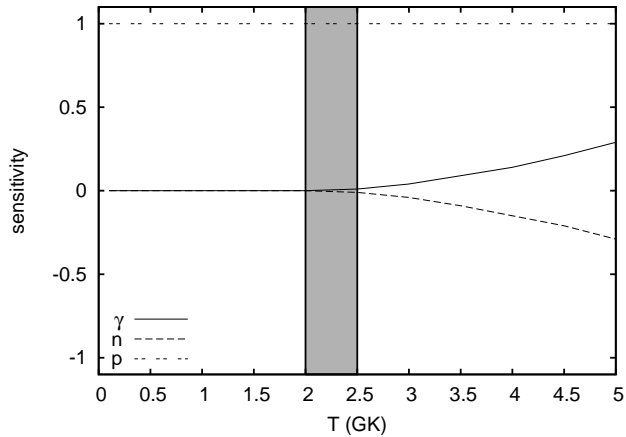}
\caption{\label{fig:sensi_rat}Sensitivity of the astrophysical reaction rates for $^{152}$Gd(p,$\gamma$)$^{153}$Tb 
to a variation of total proton-, neutron-, and $\gamma$-widths as function of plasma temperature $T$. The temperature range relevant for the nucleosynthesis of heavy $p$ nuclei is marked by the shaded area.}
\end{figure}

\begin{figure}
\includegraphics[angle=-90,width=\columnwidth]{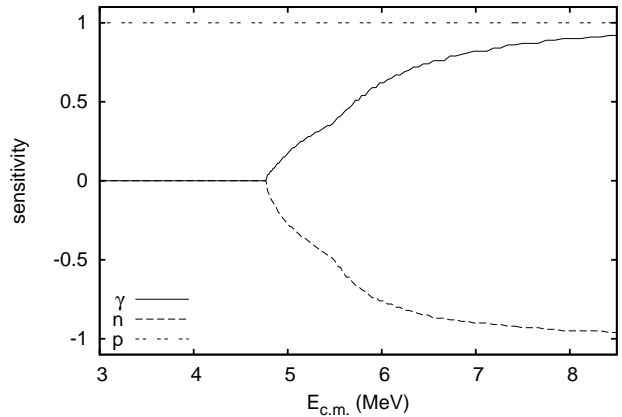}
\caption{\label{fig:sensi_pg}Sensitivity of the cross sections of $^{152}$Gd(p,$\gamma$)$^{153}$Tb 
to a variation of total proton-, neutron-, and $\gamma$-widths as function of center-of-mass energy $E_\mathrm{c.m.}$.}
\end{figure}

\begin{figure}
\includegraphics[angle=-90,width=\columnwidth]{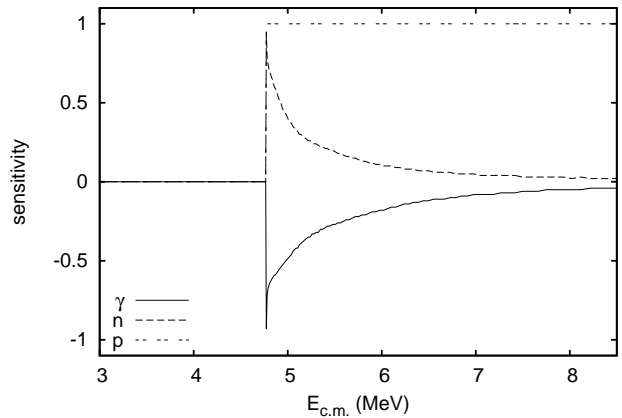}
\caption{\label{fig:sensi_pn}Same as Fig.\ \ref{fig:sensi_pg} but for $^{152}$Gd(p,n)$^{152}$Tb.}
\end{figure}

Although the $^{152}$Gd(p,$\gamma$) rate is not directly important in the $\gamma$ process, its sensitivity to a variation of the total proton-, neutron-, and $\gamma$ width is shown in Fig.\ \ref{fig:sensi_rat}. Similar results are also obtained for most (p,$\gamma$) reactions in the $p$ nuclei range below $N=82$. Due to the Coulomb barrier, the proton widths are quickly becoming smaller than the $\gamma$ widths and thus dominate the sensitivity. Therefore it is of astrophysical interest whether the present experimental data can confirm the predictions of proton widths. Figures \ref{fig:sensi_pg} and \ref{fig:sensi_pn} show the sensitivities of the reaction cross sections of $^{152}$Gd(p,$\gamma$)$^{153}$Tb and $^{152}$Gd(p,n)$^{152}$Tb, respectively, to variations of predicted total widths as function of center-of-mass energy, within the same energy range as shown in \mbox{Figs.\ \ref{fig:pg}, \ref{fig:pn}}. All cross sections are insensitive to the total $\alpha$ width as it is too small and therefore it is not shown in the figures.

Below the (p,n) threshold, the (p,$\gamma$) cross section is only sensitive to the proton width, above it the importance of $\gamma$ and neutron width quickly increase with increasing energy. The situation is reversed in the (p,n) reaction cross section which becomes less and less dependent on the $\gamma$ and neutron widths with increasing energy. In both reactions, however, the sensitivities on the two widths act oppositely. This has to be taken into account when trying to reproduce data for both reactions simultaneously. Regarding the proton width, it dominates the cross section sensitivity at the lowest measured energies in the (p,$\gamma$) reaction and at the highest measured energies in the (p,n) reaction.

It has to be noted that different nuclear properties enter the calculation of the widths (for details, see Sec.5.4 in \cite{raureview}). For the particle widths, most important are optical potentials required for the calculation of transmission coefficients and low-lying excited states. In the radiation width, transitions to states at higher excitation energy are important \cite{raugamma} and therefore this width is sensitive to the choice of $\gamma$-strength function and nuclear level density. The latter enters because it is used above the last included discrete excited state in a nucleus (see, e.g., Eq.64 in \cite{raureview}).

Close to stability, nuclear spectroscopic information is abundant but nevertheless it is not trivial to decide at which excitation energy to set the cut-off for inclusion of experimentally determined nuclear levels in each nucleus. It is essential for the correct prediction of particle widths to use a \textit{complete} level scheme. Towards higher excitation energies, more levels are missed in experimental studies and the level information given in the usual databases cannot be considered to be complete anymore. This may lead to misestimated widths in the calculation when contributing transitions are not included due to the missing excited states. In this case it is advantageous to only include the lowest experimental levels and use a theoretical level density above them, even when further levels have been identified at higher excitation energies    (see, e.g., Sec. 5.4.2 in \cite{raureview}). A practical example for this has been discussed in \cite{kiss12}. This is not an issue in $\gamma$ widths close to stability because the mainly contributing $\gamma$ transitions involve states at high excitation energies, in the unresolved resonance region, anyway \cite{raugamma}.

The theoretical calculations shown in \mbox{Sec.\ \ref{sec:theoresults}} have been obtained with version \mbox{0.8.4s} of the nuclear reaction code SMARAGD \cite{SMARAGD}, which makes use of level schemes from \cite{ENSDF}, also included in the 2010 version of NuDAT \cite{nudat}. At most 40 experimental states are used for each nucleus, as long as they are found as a consecutive sequence of levels with known spin and parity assignment. A peculiar situation arises for the neutron widths in the reactions studied here: the (p,n) energy range measured here extends into a region of incomplete level information because only rotational bands with spins \mbox{$J\geq 8$} are known above 345 keV excitation energy in $^{152}$Tb. Varying neutron- and $\gamma$-widths, as well as the included $^{152}$Tb levels, it was found that a simultaneous reproduction of the (p,n) and (p,$\gamma$) data is only possible when using a theoretical nuclear level density above \mbox{345 keV}. This indicates that further, unidentified low-spin levels must be present in addition to the experimentally determined rotational bands, which is not surprising as the total level density is expected to strongly increase with increasing excitation energy. No such problem was found for the proton width, indicating that the relevant excited states at low energy are already included in the experimental level scheme.

It should further be noted that the above treatment of excited states was also used to calculate the sensitivities shown in Figs.\ \ref{fig:sensi_rat}$-$\ref{fig:sensi_pn}. Therefore they differ from the sensitivities given in \cite{rausensi}, which used the default NuDAT set of experimental states.

\subsubsection{The optical p+$^{152}$Gd potential}
\label{sec:theoresults}

As pointed out above, the low-energy proton width is the quantity of astrophysical interest here. As shown in Figs.\ \ref{fig:sensi_pg} and \ref{fig:sensi_pn} the cross sections are sensitive to the proton widths across all investigated energies. The additional dependence on the neutron- and $\gamma$-widths can be addressed by simultaneously comparing to the (p,$\gamma$) and (p,n) data which exhibit different sensitivities, as discussed above. Since the sensitivities are almost symmetric in the neutron- and $\gamma$ widths, only the ratio between the two widths can be determined. For the reactions discussed here, however, the neutron width is already well determined through the requirement of (p,n) cross section reproduction close above the threshold, which was also used to study the excited states cut-off \mbox{(Sec.\ \ref{sec:sensi})}. Any remaining discrepancies between theoretical and experimental cross sections at the high end of the measured energy range of the (p,$\gamma$) reaction and close to the (p,n) threshold are likely due to deficiencies in either the theoretical description of the proton width through an optical potential or the $\gamma$ width, which includes two only theoretically known quantities, the $\gamma$-strength function and the nuclear level density. The action of the two widths can be distinguished, however, through their different impact on the (p,$\gamma$) and (p,n) cross sections.

\begin{figure}
\includegraphics[angle=-90,width=\columnwidth]{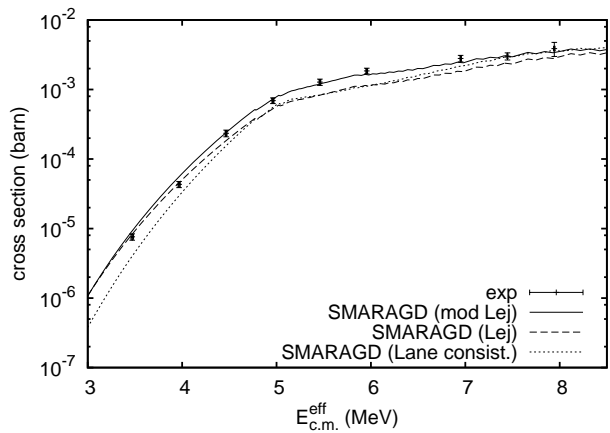}
\caption{\label{fig:pg}Experimental cross sections (exp) of $^{152}$Gd(p,$\gamma$)$^{153}$Tb 
 as function of effective center-of-mass energy $E_\mathrm{c.m.}^\mathrm{eff}$, compared to statistical model calculations using different optical p+$^{152}$Gd potentials: the potentials by \cite{jlm,lej} (Lej), by \cite{baugelane} (Lane consist.), and by \cite{Kiss07,Kiss08,Rau09} (mod Lej).}
\end{figure}

\begin{figure}
\includegraphics[angle=-90,width=\columnwidth]{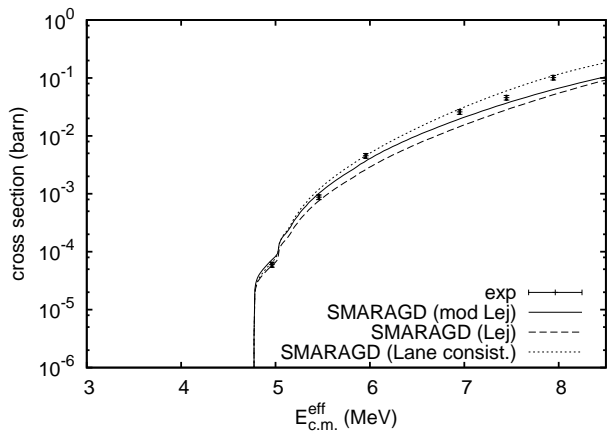}
\caption{\label{fig:pn}Same as Fig.\ \ref{fig:pg} but for $^{152}$Gd(p,n)$^{152}$Tb.}
\end{figure}

We compared the measured cross sections to SMARAGD predictions using different popular optical \mbox{p+nucleus} potentials while keeping all other ingredients to the calculations fixed. The results are shown in Figs.\ \ref{fig:pg} and \ref{fig:pn}. Essential is a simultaneous reproduction of (p,$\gamma$) and (p,n) data. Of special interest is the comparison with the three lowest (p,$\gamma$) data points which are below the (p,n) threshold, where the cross sections are only sensitive to the proton width. At all other energies, the impacts of neutron-, proton-, and $\gamma$-widths have to be disentangled. The semi-microscopic optical potential of \cite{jlm}, including low-energy modifications for astrophysics by \cite{lej} (marked 'Lej' in Figs.\ \ref{fig:pg}, \ref{fig:pn}), was also used in the large-scale reaction rate calculations of \cite{Raus00}. It results in a different energy dependence than observed in the low-energy (p,$\gamma$) cross sections although the absolute magnitude is reproduced well at these energies. There is also a problem at the higher energies. Since the predicted cross sections are below the data for both reactions, it is not possible to amend this problem by changing the $\gamma$- and/or neutron width.

Similarly to the potential by \cite{jlm,lej}, the more recent potential by \cite{baugelane} (marked 'Lane-consist.' in the figures) is the Lane-consistent version of a potential obtained from a Br\"uckner-Hartree-Fock model using a Local Density Approximation but with parameters fitted to recent experimental data. Although the (p,n) cross sections at higher energy are reproduced well with this potential, the low-energy (p,$\gamma$) cross sections are significantly below the data. Changing the ratio of the neutron- and $\gamma$-width cannot improve the overall agreement as it mostly affects the cross sections at the highest energies in the (p,$\gamma$) reaction and the ones at the lowest energies in the (p,n) reaction, while the (p,$\gamma$) cross sections below the (p,n) threshold remain unchanged.

Finally, it was found in previous work \cite{Kiss07,Kiss08,Rau09} that an improved description of low-energy (p,$\gamma$) data is possible with a modified imaginary part of the potential by \cite{lej}. Also for the two reactions discussed here, use of this potential (marked as 'mod Lej' in the figures) yields the best overall description of the data. A further slight improvement at low (p,n) and high (p,$\gamma$) energies can be achieved by increasing the $\gamma$ width by about 10\% but this does not change the conclusions regarding the proton optical potential.

\section{Summary and conclusions}
\label{sec:summary}

The reaction cross sections of $^{152}$Gd(p,$\gamma$)$^{153}$Tb 
and $^{152}$Gd(p,n)$^{152}$Tb have been measured by the
activation method at effective center-of-mass energies \mbox{$3.47
\leq E_\mathrm{c.m.}^\mathrm{eff}\leq 7.94$ MeV} and \mbox{$4.96
\leq E_\mathrm{c.m.}^\mathrm{eff} \leq 7.94$ MeV}, respectively, 
in order to extend the experimental database towards the heavier mass
region for astrophysical reactions and to test
the reliability of statistical model predictions. For the first time, (p,$\gamma$) cross sections below 
the (p,n) threshold of p+$^{152}$Gd were obtained, allowing to study 
the prediction of proton widths well below the Coulomb barrier. 

Although the cross sections depend on a number of nuclear properties, by combining the (p,$\gamma$) and (p,n) data it was possible to disentangle the contributions of these ingredients and to focus on a test of the optical p+$^{152}$Gd potential. The measured cross section values were compared to Hauser-Feshbach statistical model calculations using the nuclear code SMARAGD \cite{SMARAGD}. A good reproduction of all data across all measured energies can be obtained with the recently suggested potential by \cite{Kiss07} which is a modification of \cite{jlm,lej}.

Further experiments to obtain (p,$\gamma$) data at even lower energies than studied here would be desirable but will prove very challenging due to the tiny reaction cross sections.

\begin{acknowledgments}
This work was supported by the ESF EUROCORES program EUROGENESIS and by the Scientific and Technological Research
Council of Turkey TUBITAK-Grant-109T585, OTKA (K101328 and K108459). TR acknowledges support by the BRIDGCE grant from the UK Science and Technology Facilities Council (grant ST/M000958/1), by the Swiss National Science Foundation, and the European Research Council (grant GA 321263-FISH).
\end{acknowledgments}

\end{document}